\documentclass[sigconf]{acmart}
\usepackage{booktabs}
\usepackage{enumitem}
\usepackage{tabularx}
\usepackage{graphicx}
\usepackage{subcaption}

\acmConference[]{}{}{}
\acmYear{}
\copyrightyear{}
\acmDOI{}
\acmISBN{}

\AtBeginDocument{%
  \providecommand\BibTeX{{%
    \normalfont B\kern-0.5em{\scshape i\kern-0.25em b}\kern-0.8em\TeX}}}

\begin{document}
\title{Network Traffic Analysis based IoT Device Identification}

\author{Rajarshi Roy Chowdhury}
\affiliation{\institution{Faculty of Integrated Technologies}}
\email{19h0901@ubd.edu.bn}

\author{Sandhya Aneja}
\affiliation{\institution{Faculty of Integrated Technologies}}
\email{sandhya.aneja@ubd.edu.bn}

\author{Nagender Aneja}
\affiliation{\institution{Institute of Applied Data Analytics}}
\email{nagender.aneja@ubd.edu.bn}

\author{Emeroylariffion Abas}
\affiliation{\institution{Faculty of Integrated Technologies}}
\email{emeroylariffion.abas@ubd.edu.bn}

\begin{abstract}
Device identification is the process of identifying  a device on Internet without using its assigned network or other credentials. The sharp rise of usage in Internet of Things (IoT) devices has imposed new challenges in device identification due to a wide variety of devices, protocols and control interfaces. In a network, conventional IoT devices identify each other by utilizing IP or MAC addresses, which are prone to spoofing. Moreover, IoT devices are low power devices with minimal embedded security solution.  To mitigate the issue in IoT devices, fingerprint (DFP) for device identification can be used. DFP identifies a device by using implicit identifiers, such as network traffic (or packets), radio signal, which a device used for its communication over the network. These identifiers are closely related to the device hardware and software features. In this paper, we exploit TCP/IP packet header features to create a device fingerprint utilizing device originated network packets. We present a set of three metrics which separate some features from a packet which contribute actively for device identification. To evaluate our approach, we used publicly accessible two datasets. We observed the accuracy of device genre classification 99.37\% and 83.35\% of accuracy in the identification of an individual device from IoT Sentinel dataset. However, using UNSW dataset device type identification accuracy reached up to 97.78\%.
\end{abstract}



\maketitle
\pagestyle{plain}  
\pagestyle{empty}  

\section{Introduction}
Device identification is the process of identifying a device on Internet without using its network identifiers or other credentials such as IP address, Medium Access Control (MAC) address, Electronic Serial Number (ESN), International Mobile Station Equipment Identity (IMEI) number or Mobile Identification Number (MIN). These identifiers can be altered or manipulated by using expert knowledge of networking or software, which can create significant security threats where devices identification play key security roles. DFP exploits device-specific signature or packet information (feature set) which the device uses for communication over the network \cite{aneja2018iot}.  Effective device fingerprint must assure two attributes (i) the features are hard to forge, (ii) the DFP remains stable even when devices move from one network to another network \cite{xu2015device}. DFP has emerged as a significant solution for device identification (or authentication) due to its resistance against vulnerabilities such as node forgery or masquerading in IoT network. Features used for DFP are classified based on their extraction methods. There are two techniques namely active and passive which have been observed to impact the identification of devices. In active fingerprinting, devices are probed with different types of packets and the subsequent received responses are analyzed to attain unique fingerprints. In contrast, in passive fingerprinting, a profiler observes ongoing communication of the target system without any obvious querying into the system, by analyzing and extracting information from transmitted packets. In this paper, we used passively observed datasets to evaluate the proposed method, our DFP approach is for passive fingerprinting.

A DFP approach proposed based on the analysis of (each packet) 23 features from 12 network packets by Miettinen et al. \cite{miettinen2017iot}. The set of features were represented as binary (0 or 1) and integer values, where binary value 1 define the protocol used during communication. Overall, the scheme achieved identification accuracy of 81.50\% (global ratio) over the 27 devices using the IoT Sentinel dataset \cite{miettinen2017iot}. 

The single packet-based DFP method  \cite{aksoy2019automated} uses genetic algorithm (GA) to determine the feature subset which contributes substantially, for generation of device fingerprint used for device identification. This approach has been shown to achieve approximate classification accuracy of over 95\% for device genre and 82\% for individual device type using 23 IoT devices from IoT Sentinel dataset \cite{miettinen2017iot}. 

We worked on a device fingerprinting approach using packets header information (distinct protocols header features) from the target devices to extract a unique set of features (actual feature values) to generate device fingerprint. Features are assessed using three metrics: the variability of a feature, the stability of a feature, and the suitability of a feature, to learn the most suitable feature subset. Metrics evaluation results assign a score to each of the features to form a feature vector. The feature set is subsequently processed to filter feature subset according to the user-defined threshold value $\lambda$ whereas a lambda value ranged between 0 and less than 1. Finally, we used machine learning (ML) classification algorithms to classify individual IoT device type or genre by using selected feature subset vectors of each device. While two algorithms, such as J48 and PART, outperformed compare to other classification algorithms. However, based on the user expected accuracy $\lambda$ value can be changed empirically. To evaluate our proposed DFP method we used two publicly available datasets: IoT Sentinel dataset  \cite{miettinen2017iot} and UNSW dataset  \cite{sivanathan2018classifying}. While we achieved approximate classification accuracy to 99.37\% for device genre and 83.35\% for individual device type by using IoT Sentinel dataset \cite{miettinen2017iot} and 97.78\% of precision achieved using 10678 instances from UNSW dataset \cite{sivanathan2018classifying} for identification of individual device type. The contributions of this paper are the following:

\begin{itemize}
	\item We present a conceptual IoT network model incorporating device identification with a wide range of IoT devices (Section 2).  
	\item We introduce metric-entropy based packet header features assessment methodology using three metrics for device fingerprinting and design of our proposed DFP approach (Section 5).
	\item We examine proposed DFP method using two datasets, which consists of various real-world IoT devices (Section 6).
\end{itemize}

The rest of this paper is organized as follows: Section 2 describes a conceptual IoT network model to depict the DFP incorporated in the IoT network. In Section 3, notations and device information are presented, and preliminary of the metrics evaluation process is discussed in Section 4. The proposed DFP approach is illustrated in Section 5, while experimental results with the different datasets are discussed in Section 6. Section 7 addresses related work and Section 8 presents the conclusion and the future direction of work. 

\section{IoT Network Model incorporating Device Identification}
We presume that our network model comprises of various types of IoT devices, which are used in smart homes, smart buildings or enterprises \cite{shahid2017internet}. In a network, most of the devices, such as IP cameras (TP-LinkCam, EdimaxCam), smart bulbs, connect to the user's network (a single access point gateway) directly either through WiFi (wireless fidelity) or Ethernet connection to access the Internet services. 

Most of the devices are identified by using explicit identifiers, such as IP address, MAC address and other network identities. Devices use these identifiers to communicate on the network. Unfortunately, all these identifiers have been exhibited to be easily mutable by using software \cite{douceur2002sybil}. Spoofing is one of the techniques which can forge a device explicit identifier to gain illegitimate access to restricted resources in the network. For example, an ioctl system call can be used to forge or modify MAC address of a device network interface card (NIC) \cite{xu2015device} to access a network (e.g. a corporate office) where devices are recognized by using their MAC addresses. Similarly, IoT Node using spoofed IP address can launch crucial attacks while communicating with an access point (AP) to get required services. Initially, each device on the IoT network is identified by using their unique IPv4 address, however, even IP address is spoofed than also it would be unique. Moreover the range of spoofed IP address is restricted by type of the network e.g. 256 in case of Type C network.

\begin{figure}[h]
	\centering
	\includegraphics[width=\linewidth, height=5.5cm]{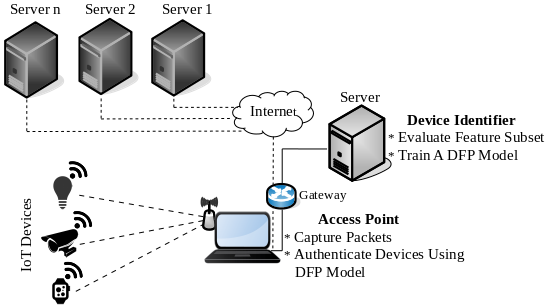}
	\caption{Conceptual IoT network model of the proposed method}
	\label{fig:iot_network_model}
\end{figure}

IoT devices are classified as resource-constrain devices \cite{InternetDraft} in terms of processing power, communication capability, memory and energy, while complex cryptography algorithms can be used to secure the device identity. Therefore, while  communicating  the encrypted data, the processing power of devices must be preserved. In other words, a secure method is required to identify devices without using the traditional identifiers, such as IP address or MAC address. An intuition that network traffic provides device-specific signature since devices comprised different types of hardware-software and used distinct communication protocols in a network. DFP is a technique to identify IoT devices using network traffic information from the target devices. Conceptual network model is shown in Figure \ref{fig:iot_network_model} wherein the proposed DFP method can be used to generate unique identifiers of IoT devices to overcome the spoofed network identities of the device.

We assume that our proposed DFP scheme is able to capture all types of network traffic (inbound and outbound traffic) passively on the access point (AP). The IoT Node communicate for various services e.g. DNS to translate domain name of Internet resources, SSL/TLS certificates to encrypt and decrypt data, NTP server to synchronize device time, and Manufacture server to upgrade firmware services. This communication of IoT Node is captured on the AP shown in Figure \ref{fig:iot_network_model}. In the proposed model packets payload information are not considered, which can be plain text, compressed, binary or encrypted data. We only focused on packet header features (or fields) \cite{bezawada2018iotsense}, \cite{miettinen2017iot}, \cite{aksoy2019automated}. An intuition that distinct protocols features from the packet header provide a significant fingerprint for device identification. This fingerprint is generated using header information of all the packets and regularly updated on the device identifier station. Device identifier evaluates the feature set from \textit{n} number of packets from each device to determine the most suitable subset of features and then train a model for devices classification using an ML classification algorithm. Finally, the AP uses this classification model to authenticate devices in the IoT network.

\section{Notations and Datasets}
A list of notations used to describe our proposed DFP approach are listed in Table \ref{tab:notation} along with a description of each notation.\\

\begin{table}[h]
	 \caption{A list of notations}
	\label{tab:notation}
	\begin{tabular}{cl}
		\toprule
		Notation & Description \\
		\midrule
		$pkt$ & : A TCP/IP packet\\
		$t$ & : Total number of packets or instances\\
		$r$ & : $r^{th}$ number of packet\\
		$f$ & : A feature in a packet\\
		$m$ & : Total  number of features in a packet\\ 
		$j$ & : $j^{th}$ number of feature in a packet\\
		$d$ & : An IoT Device\\
		$n$ & : Total number of devices in a dataset\\ 
		$blk$ & : A block of packets\\
		$k$ & : Total number of packets in a block\\ 
		$b$ & : Total number of blocks\\
		$str$ & : A string (concatenated features values)\\
		$c$ & : Total number of distinct characters in a string\\
		$i$ & : $i^{th}$ number of character in a string\\
		$P(str_{f_{ji}})$ & : Discrete probability density function\\
		$H(str_{f_j})$ & : Metric-entropy value calculation function\\
		$L$ & : Total length of a string\\
		$sd$ & : Individual device stability vector\\
		$v$ & : Variability vector of the features\\
		$s$ & : Stability vector of the features \\
		$u$ & : Suitability vector of the features\\
		$l$ & : $l^{th}$ number of device in a dataset\\
		$\|$ & : Concatenation operation\\
		$\lambda$ & : A threshold value (Lambda)\\
		$\ast$ & : Multiplication operation\\
		\bottomrule
	\end{tabular}
\end{table}

We utilized publicly available two datasets to evaluate our proposed DFP method. The datasets are IoT Sentinel \cite{miettinen2017iot} dataset and UNSW \cite{sivanathan2018classifying} dataset containing different types of IoT devices. We use a total of 27 IoT devices from IoT Sentinel dataset. There are a total 12 genres or groups of devices in IoT Sentinel \cite{miettinen2017iot}: Aria, D-Link (D-LinkCam, D-LinkDayCam, D-LinkDoorSensor, D-LinkHomeHub, D-LinkSensor, D-LinkSiren, D-LinkSwitch and D-LinkWaterSensor), Hue (HueSwitch and HueBridge), Smarter (iKettle2 and SmarterCoffee), Lightify, MAXGateway, TP-Link (TP-LinkPlugHS100 and TP-LinkPlugHS110), WeMo (WeMoInsightSwitch, WeMoLink and WeMoSwitch) and Withings. We use a total of 19 devices from UNSW dataset. All these devices information are listed in Table \ref{tab:iot_sentinel} and Table \ref{tab:unsw}.

\begin{table}[h]
\caption{IoT Sentinel dataset- the list of devices \cite{miettinen2017iot}}
\label{tab:iot_sentinel}
\begin{tabular}{cll}
	\toprule
	Manufacturer & Device & Connectivity\\
	\midrule
	Fitbit & Aria & WiFi \\
	D-Link & D-LinkCam & WiFi \\
	$\star$ & D-LinkDayCam  & WiFi/Ethernet\\
	$\star$ & D-LinkDoorSensor & Z-Wave\\
	$\star$ & D-LinkSensor & WiFi\\
	$\star$ & D-LinkWaterSensor & WiFi\\
	$\star$ & D-LinkSiren & WiFi\\
	$\star$ & D-LinkHomeHub & WiFi/Ethernet/Z-wave\\
	$\star$ & D-LinkSwitch & WiFi\\
	Edimax & EdimaxCam1 & WiFi/Ethernet\\
	$\star$ & EdimaxPlug1101W & WiFi\\
	$\star$ & EdimaxPlug2101W & WiFi\\
	Ednet & EdnetCam1 & WiFi/Ethernet\\
	Ednet.Living & EdnetGateway & WiFi/Other\\
	HomeMatic & HomeMaticPlug & Other\\
	Philips Hue & HueBridge & ZigBee/Ethernet\\
	$\star$ & HueSwitch & ZigBee\\
	Smarter & iKettle2 & WiFi\\
	$\star$ & SmarterCoffee & WiFi\\
	Osram  & Lightify & WiFi/ZigBee\\
	eQ-3 & MAXGateway & Ethernet/Other\\
	TP-Link & TP-LinkPlugHS100 & WiFi \\
	$\star$ & TP-LinkPlugHS110 & WiFi\\
	Belkin (Wemo) & WeMoInsightSwitch & WiFi\\
	$\star$ & WeMoLink & WiFi/ZigBee\\
	$\star$ & WeMoSwitch & WiFi\\
	Withings & Withings & WiFi\\
	\bottomrule
\end{tabular}
\end{table}

\begin{table}[h]
	\caption{UNSW dataset - the list of devices \cite{sivanathan2018classifying}}
	\label{tab:unsw}
	\begin{tabular}{cll}
		\toprule
		Manufacturer & Device & Connectivity\\
		\midrule
		Amazone & AmazoneEcho & WiFi \\
		Belkin & BelkinMotion & WiFi \\
		$\star$ & BelkinSwitch & WiFi \\
		Blipcare & BlipcareBpMeter & WiFi \\
		Google Nest & DropCam & WiFi \\
		HP & HP-Printer & WiFi \\
		iHome & iHomePlug & WiFi \\
		Lifx & LiFX & WiFi \\
		Nest & NestSmoke & WiFi \\
		Netatmo & NetatmoCam & WiFi \\
		$\star$ & NetatmoWeather & WiFi \\
		Pixstar & PixstarPhotoFrame & WiFi \\
		Samsung & SamsungCam & WiFi \\
		SmartThings & SmartThings & Wired \\
		TP-Link & TP-LinkCam & WiFi \\
		$\star$ & TP-LinkPlug & WiFi \\
		Triby & TribySpeaker & WiFi\\
		Withings & WithingsScale & WiFi\\
		$\star$ & WithingsSleep & WiFi\\
		\bottomrule
	\end{tabular}
\end{table}

[Note: $\star$ indicates devices are from the same manufacturer as above]

\section{ Preliminaries}
Robyns et al. \cite{robyns2017noncooperative} defined three metrics: (a) the variability of a bit (b) the stability of a bit and (c) the suitability of a bit for bitwise entropy evaluation of a Probe-Request frame on a WiFi network.

The defined metric variability attains ideal value $1$ when the feature (particular field of packet format) over the multiple transmission from the devices is highly variable over all the devices and therefore uniquely contribute to fingerprint of the device e.g IP address of a device is unique and therefore contribute highly to the feature set for fingerprinting the device.  
	
Similarly, the defined metric stability attains ideal value $1$ when the feature among other features of the device uniquely identify the device and remains stable over the multiple transmissions from the device. The defined metric suitability uses variability, and stability by multiplying to calculate the feature suitability. Robyns et al. evaluated suitability of features e.g. SSID, vendor specific data, HT capabilities in a frame to identify the mobile devices. The authors included the MAC address while evaluating frame for the metrics.

We define the metrics at character-level for packet header analysis on 212 set of features and improve the IoT Sentinel dataset classification compared to \cite{aksoy2019automated} without IP address and the method outperformed with IP address.  The character-level metric is more promising, as the bit-wise metric evaluate the probability density of frame vector over $\{0, 1, none\}$ while character level metric evaluate the probability density value of feature vector over$\{0, 1,\ldots , 9, none\}$. The bit wise evaluation \cite{robyns2017noncooperative} drops the accuracy gradually e.g the reported accuracy of 10-100 devices including MAC Address is 60-87\%.

\section{ Device Identification- Device Fingerprinting Approach}
\begin{figure}[h]
	\centering
	\includegraphics[width=\linewidth, height=10.5cm]{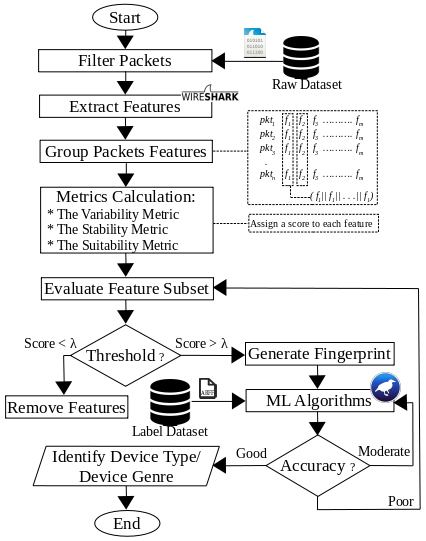}
	\caption{A workflow of the proposed DFP method}
	\label{fig:flowchart}
\end{figure}

In this section, we describe major components of our device fingerprinting method to identify IoT device type or genre based on a single TCP/IP packet. This method appraises Network layer (IP and ICMP), Transport layer (TCP and UDP) and Application layer (DNS, HTTP, TLS/SSL and DHCP) protocols header fields (or features) to learn the most suitable subset of features among a large number of features, which can be used for device fingerprinting. We define three character-level metrics, variability of a feature, stability of a feature, and suitability of a feature to assess the feature set. This process identifies a suitable subset of features which are unique to each device, while a lambda value is utilized to filter selected feature subset according to the expected accuracy. Then we perform a machine learning classification algorithm to identify or classify individual IoT device type or genre by using a selected subset of features. The entire process of our proposed DFP scheme is illustrated in Figure \ref{fig:flowchart}.

\subsection{Packet Header Features}
In the proposed system, we used device originated network traffic or packets to extract a list of features from each packet header fields. A total $m$ = 218 features extract from a single packet $pkt_r = (f_1, f_2, f_3, f_4, f_5, f_6, \ldots \ldots , f_{m-1}, f_m)$ based on the contents of a packet header information, where $r \in$ ($1, 2, 3 \ldots \ldots, t-1, t$). Most of the values of the features are of integer type except from some features from the DHCP protocol, which include both integer and string values. For a range of string values, we convert to nominal values by using $StringToNominal$ function in WEKA tool \cite{Weka}. Additionally, we eschew IP address and IP checksum for the proposed DFP. The set of features used for our DFP scheme is partially listed in Table \ref{tab:feature_set}, where all features are based on distinct protocols header fields to ensure that the proposed method can be used for all types of network traffic like as plain-text, compressed data, binary or encrypted traffic.

\begin{table}[h]
	\caption{List of the protocols header fields (or features)}
	\label{tab:feature_set}
	\begin{tabular}{lll}
		\toprule
		Layers & Protocols & Features\\
		\midrule
		Link Layer & - - - & - - -  \\
		\hline 
		Network Layer & IP, ICMP & ip.hdr\_len, \\
		 & (33 features) & ip.dsfield, \\
		 & & ip.dsfield.dscp, \\
		 & & ip.proto, ip.ttl, \\
		 & & ip.opt.len, ip.opt.ra, \\
		 & & ip.opt.type, icmp.code, \\
		 & & icmp.checksum,\\
		 & & icmp.type, icmp.seq, etc.\\
         \hline
		 Transport Layer & TCP, UDP   & tcp.srcport,\\
		 &(64 features)& tcp.dstport, tcp.hdr\_len,\\
	     & & tcp.window\_size,\\
		 & & tcp.checksum,\\
		 & & tcp.connection.syn,\\
		 & & tcp.analysis.ack\_rtt,\\
		 & & udp.srcport,\\
		 & & udp.checksum,\\
		 & & udp.checksum.status,\\
		 & & udp.stream, etc.\\
		 \hline
		 Application Layer & DNS, HTTP, &  dns.flags, dns.id,\\
		 & SSL/TSL, & dns.qry.name.len,\\
		 & DHCP & dns.count.queries,\\
	 	 & (121 features) & http.request,\\
	 	 & & http.response,\\
	 	 & & http.request\_number,\\
	 	 & & tls.handshake.version,\\
	 	 & & tls.handshake, dhcp.id,\\
	 	 & & dhcp.option.hostname, etc.\\
	    \bottomrule
	\end{tabular}
\end{table}

\subsection{Feature Subset Selection}
We noticed that from our literature studies  \cite{aksoy2019automated}, \cite{bezawada2018iotsense}, \cite{ammar2019network}, \cite{radhakrishnan2014gtid}, \cite{pinheiro2019identifying} a small number of features is sufficient to identify most of the devices with high accuracy. Therefore, instead of manually or randomly selecting a suitable subset of features for DFP, it would be rational to identify the most suitable feature subset automatically from a large number of protocol header features. To accomplish this goal our proposed DFP methodology comprises of the character-level analysis on packet header features using the three metrics \cite{robyns2017noncooperative}: (a) the variability metric of a feature (b) the stability metric of a feature and (c) the suitability metric of a feature.

We measure on three character-level metrics of a feature obtained from $b$ blocks of packets transmitted by the device. Hence, from every device $b$ number of blocks values are concatenated individually based on each feature, such as $blk_1(f_j) \| blk_2( f_j) \| \ldots \| blk_{b-1}(f_j) \| blk_b(f_j)$, $j \in$ ($1, 2, 3, \ldots \ldots, m-1, m$), to determine a metric-entropy \cite{shannon1948mathematical} value of that particular feature, wherein, a block comprises of $k$ number of packets transmitted by an IoT device over the network during communication. 

A feature values from $k$ number of packets are concatenated, like as $pkt_1(f_j) \| pkt_2(f_j) \| .. \| pkt_{k-1}(f_j) \| pkt_k(f_j)$, to form a string $str_{f_j}$ to compute metric-entropy (Eq. \ref{metric_entropy}) value of that feature.

\begin{equation}
\label{metric_entropy}
H(str_{f_j})= \frac{- \sum_{i=1}^{c}P(str_{f_{ji}})\times log_2 P(str_{f_{ji}})}{L}
\end{equation}

Where metric-entropy defines randomness of a feature string, $P(str_{f_{ji}})$ characterizes discrete probability density of individual character from a given string and a total number of unique character in a string is delimited by $c$. Each metric-entropy value ranges from 0 to less than 1, wherein 0 means no entropy value.

\textbf {Calculation of The Variability Metric:} A feature variability is assessed based on the metric-entropy value of that feature calculated from a different set of devices, where equal number of blocks of packets were chosen from each device for assessment. A feature metric-entropy value calculates from $n$ number of blocks, while blocks are concatenated, such as $d_1blk_1str_{f_1} \| d_2blk_1str_{f_1} \| d_3blk_1str_{f_1} \\ \| \ldots  \| d_nblk_1str_{f_1}$, $d_1blk_1str_{f_2} \| d_2blk_1str_{f_2} \| d_3blk_1str_{f_2} \| \ldots \ldots \ldots \| \\d_nblk_1str_{f_2}$, \ldots, $d_1blk_1str_{f_m} \| d_2blk_1str_{f_m} \| d_3blk_1str_{f_m} \| \ldots \ldots \ldots \| \\d_nblk_1str_{f_m}$, individually to compute variability (Eq. \ref{metric_entropy}) value of that feature. The variability vector $v$ can be demonstrated as: $$v = [H_{str_{f_1}}, H_{str_{f_2}}, H_{str_{f_3}}, H_{str_{f_4}}, H_{str_{f_5}}\ldots, H_{str_{f_m}}]$$   

The objective of this variability metric is to identify the least number of features which provide a unique contribution for device fingerprinting.

\textbf {Calculation of The Stability Metric:} 	  	  
The stability vector $s$ define how likely feature values remain stable over multiple transmission from the same device, which can be used for device fingerprinting. Consequently, first we measure individual device stability vector $sd_l$ represented as: $$sd_l = [1 - H_{str_{f_1}}, 1 - H_{str_{f_2}}, 1 - H_{str_{f_3}} \dots, 1 - H_{str_{f_m}}]$$ 
Then, device-specific stability vector from all devices are averaged to assess the final stability vector $s$. The goal of the stability metric is to identify a set of features that are associated with devices and varies less frequently.

\begin{equation}
\label{average_statbility}
s(f_j)= \frac{\sum_{l=1}^{n}sd_l(f_j)}{n}
\end{equation}

\textbf {Calculation of The Suitability Metric:} We determine the most suitable set of features by combining the variability and stability metrics. Hypothetically, the features utilized for our DFP approach should be highly variable and stable to achieve higher classification accuracy. For instance, a feature metric-entropy value with $0.5$ variability and $0.99$ stability provides approximately higher suitability, whereas a feature with zero variability and $0.99$ stability is considered unsuitable to use for fingerprint. We incorporate these two metrics by using multiplication operation to attain the suitability vector $u$ of the features (Eq. \ref{statbility}): 

\begin{equation}
\label{statbility}
u = v\ast s
\end{equation}

The suitability vector $u$ can be represented as $u = [v_{f_1}\ast s_{f_1}, v_{f_2}\ast s_{f_2}, v_{f_3}\ast s_{f_3}, v_{f_4}\ast s_{f_4}, \ldots, v_{f_{m-1}}\ast s_{f_{m-1}}, v_{f_m}\ast s_{f_m}]$. Henceforth, a threshold value $\lambda$ is used to experimentally assess the feature subset according to the user's expected accuracy and to include those features which have suitability value greater than the threshold. A $\lambda$ value influences the trade-off between a feature uniqueness and suitability to be used for device fingerprint. Table \ref{tab:feature_set_SVU} presents assessment results of some features evaluated by using the three metrics, while features are randomly selected from the IoT Sentinel dataset and a $\lambda$ value defined experimentally to exclude (x) or include (\checkmark) features for device fingerprinting.

\begin{table}[h]
	\caption{Randomly selected few features suitability metric evaluation results}
	\label{tab:feature_set_SVU}
	\begin{tabular}{lll}
		\toprule
		Features & u & $\lambda$\\
		\midrule
		&  u > $\lambda$ & $0.0$ \\
		\hline
		ip.hdr\_len &  0.0002 & \checkmark  \\
		ip.dsfield &  0.0 & x  \\
		ip.dsfield.dscp &  0.0001 & \checkmark \\
		ip.proto &  0.0005 & \checkmark \\
		ip.ttl & 0.0004 & \checkmark\\
		ip.opt.len & 0.0 & x \\
		ip.opt.ra &  0.0 & x\\
		ip.opt.type &  0.005 & \checkmark \\
		icmp.checksum &  0.0427 & \checkmark \\
		icmp.type &  0.0086 & \checkmark \\
		icmp.seq &  0.2114 & \checkmark \\
		icmp.code &  0.0 & x \\
		tcp.srcport &  0.0005 & \checkmark \\
		tcp.dstport &  0.0006 & \checkmark \\
		tcp.hdr\_len &  0.0006 & \checkmark \\
		tcp.window\_size & 0.0005 & \checkmark \\
		tcp.checksum &  0.0002 & \checkmark \\
		tcp.connection.syn & 0.0 & x \\
		tcp.analysis.ack\_rtt &  0.0 & x \\
		udp.srcport &  0.0008 & \checkmark \\
		udp.checksum & 0.0003 & \checkmark \\
		udp.stream &  0.0022 & \checkmark \\
		dns.flags & 0.0003 & \checkmark \\
		dns.id & 0.0003 & \checkmark \\
		dns.qry.name.len1 & 0.0066 & \checkmark \\
		dns.count.queries & 0.0052 & \checkmark \\
		http.request &  0.0 & x \\
		http.response\_number &  0.0162 & \checkmark \\
		http.chunk\_size1 & 0.1049 & x \\
		tls.handshake.version &  0.0045 & \checkmark \\
		tls.handshake3 & 0.0 & x\\
		dhcp.id & 0.0017 & \checkmark \\
	
		\bottomrule
	\end{tabular}
\end{table}

\subsection{Device Classification}
In the preceding subsection, we explained how to choose the most suitable subset of features from packet header fields that can be used for devices classification by using ML algorithms. We used WEKA (Waikato Environment for Knowledge Analysis) tool \cite{Weka} to perform various ML classification algorithms, such as PART and J48 Decision Trees, to classify IoT devices based on distinct protocol header features from a single packet of a device. We observed that PART and J48 algorithms provide higher accuracy than other algorithms available in WEKA libraries. Our proposed method utilizes real values of the features to facilitate both algorithms to identify devices with high accuracy.   

\subsubsection{Identification of device genre and individual device type}
\begin{itemize}	
	\item \textbf{Identification of device genre}: We use a total of 27 IoT devices from IoT Sentinel dataset, those devices are categorized based on their manufacturer name or device name (if a single device from manufacture) and then apply proposed DFP method to classify individual device genre. 	
	
	\item \textbf{Identification of individual device type} (based on device name): In this scenario, we utilize both datasets to identify specific device type using our proposed DFP approach.  
\end{itemize}  

\section{Experimental Results}
In this section, we describe our experimental process and results based on the two datasets. First, we extracted all the devices originated network packets or instances from the individual device (.pcap files) by using T-shark \cite{Gerald1998}, with each device identified based on their MAC address or IP address. Then, all the device instances were filtered according to the selected protocols used for our DFP approach and were also labeled according to the individual device type and genre.

\begin{figure}[h]
	\centering
	\includegraphics[width=\linewidth]{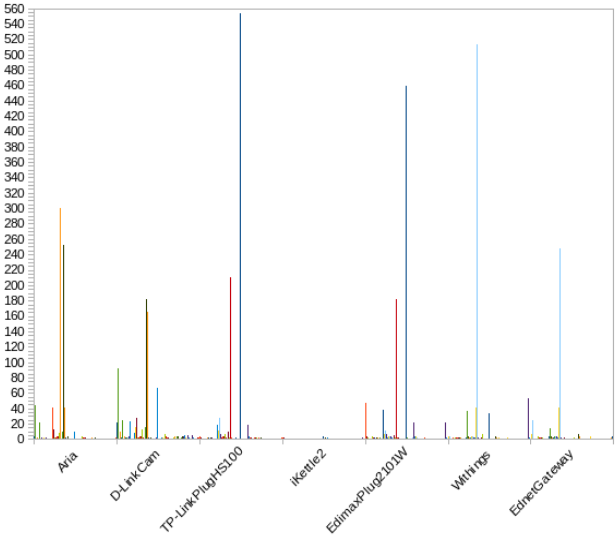}
	\caption{Individual device total number of distinct values of a feature}
	\label{fig:indiv_feat_variation}
\end{figure}

All the devices information (46 IoT devices) from both datasets are listed in Table \ref{tab:iot_sentinel} and Table \ref{tab:unsw}, on which our proposed DFP approach was carried out. In our experiment to evaluate the features, we used approximately 100 packets from each device to compute the three metrics explained in Section 5. 

However, from our observation, we noticed that in most of the cases, the total number of distinct values of a feature varies from 0 to 100, as shown in Figure \ref{fig:indiv_feat_variation} e.g. IP address has one distinct value for each device. Hence, we do not consider a feature with a large number of distinct values in our metrics computation separately. We assume that these types of features always provide a prominent result for DFP.

\begin{table}[h]
	\caption{Total number of instances (or packets) used for training and validation}
	\label{tab:training_validation}
	\begin{tabular}{lllll}
		\toprule
		Dataset & Devices & Training & Validation & Total\\
		\midrule
		& & 70\% & 30\% & 100\% \\
		\hline
		IoT Sentinel & 27 & 69489 & 29782 & 99271 \\
	    UNSW & 19 & 7474 & 3204 & 10678 \\
		\bottomrule
	\end{tabular}
\end{table}

\subsection{Classification Performance}
Every dataset instances (or packets) were randomly divided into a training dataset (70\% of the dataset) and a validation dataset (30\% of the dataset) by using WEKA tool to perform device classification or identification based on ML algorithms. We used unsupervised-resample function to split the datasets accordingly. Initially, we select the suitable feature subset from the attributes list in WEKA preprocess stage based on the metrics evaluation results and a threshold value calculated from the dataset to remove insignificant features. We performed metrics evaluation process two thousand times with different sets of packets and finally averaged the suitability metric values to assign a score for each feature. Table \ref{tab:training_validation} lists the total number of instances used for our DFP approach according to different datasets and Table \ref{tab:selected_features} reports the total number of features selected according to the threshold values in our experiment. From our observation, we notice that when the threshold value decreased devices identification accuracy and the number of features increased respect to the individual dataset.

\begin{table}[h]
	\caption{Total number of selected features based on $\lambda$ value}
	\label{tab:selected_features}
	\begin{tabular}{llll}
		\toprule
		Dataset & Total Number  & Threshold & Selected Features\\
		 & of Features & ($\lambda$) & \\
		\midrule
	    IoT Sentinel & 218 & 0.0 & 161 \\
        UNSW & 218 & 0.0004 & 86\\  
		\bottomrule
	\end{tabular}
\end{table}

\begin{figure}[h]
	\centering
	\includegraphics[width=\linewidth]{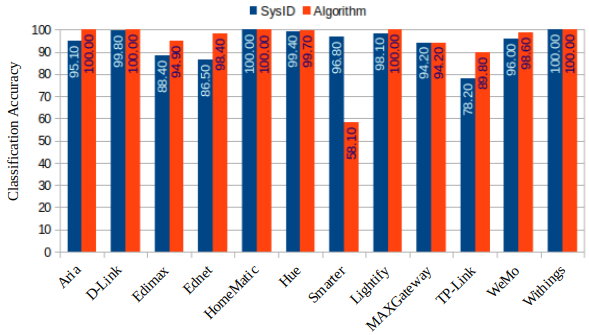}
	\caption{IoT Sentinel dataset: device genre classification performance}
	\label{fig:devices_genre}
\end{figure}

To assess the proposed DFP method classification performance, we used several ML classification algorithms to identify individual device genre and device type based on a single packet from each device. By using IoT Sentinel dataset  \cite{miettinen2017iot} our method achieved mean classification accuracy of 99.37\% for device genre with threshold value 0.0 experimentally, as shown in Figure \ref{fig:devices_genre}. There were total 12 genres or groups of devices namely Aria, D-Link (D-LinkCam, D-LinkDayCam, D-LinkDoorSensor, D-LinkHomeHub, D-LinkSensor, D-LinkSiren, D-LinkSwitch and D-LinkWaterSensor), Hue (HueSwitch and HueBridge), Smarter (iKettle2 and SmarterCoffee), Lightify, MAXGateway, TP-Link (TP-LinkPlugHS100 and TP-LinkPlugHS110), WeMo (WeMoInsightSwitch, WeMoLink and WeMoSwitch) and Withings. We grouped the devices based on their manufacture name or device name if a single device from the manufacture in the dataset.

\begin{figure*}[ht]
	\centering
	\begin{subfigure}[b]{0.45\textwidth}
		\includegraphics[width=7cm, height=4.5cm]{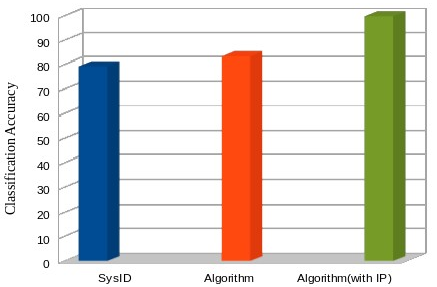}
		\caption{IoT Sentinel dataset}
		\label{fig:devices_27}
	\end{subfigure}
	~ ~ ~ 
	\begin{subfigure}[b]{0.45\textwidth}
		\includegraphics[width=7cm, height=4.5cm]{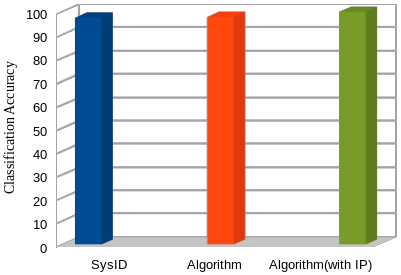}
		\caption{UNSW dataset}
		\label{fig:devices_19}
	\end{subfigure}
	\caption{Overall classification performance: (a) IoT Sentinel dataset and (b) UNSW dataset}
	\label{fig:device_classification_performance}
\end{figure*}

Figure \ref{fig:devices_genre} presents our proposed DFP scheme attained higher classification accuracy compared with SysID. However, the Smarter group of devices exhibited lower classification performance by using the proposed method. However, overall the proposed method reached a higher classification rate to identify individual manufacturer with multiple devices. 

\begin{figure*}[h]
	\centering
	\includegraphics[width=\linewidth]{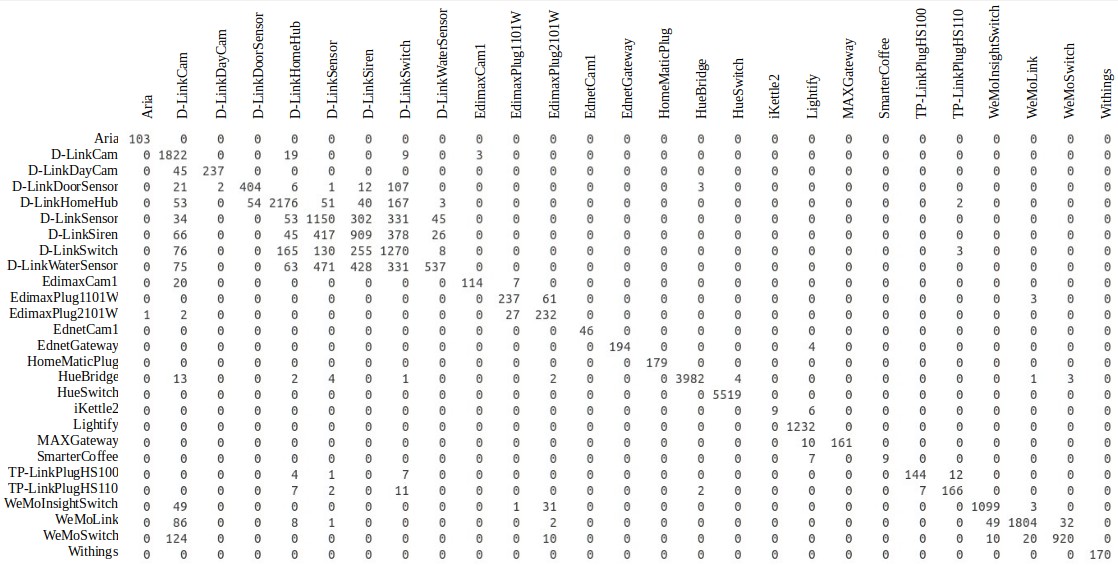}
	\caption{IoT Sentinel dataset: confusion matrix}
	\label{fig:confusoin_matrix}
\end{figure*}

In terms of individual device type classification performance, the proposed scheme achieved overall 83.35\% of accuracy using IoT Sentinel dataset \cite{miettinen2017iot} as shown in Figure \ref{fig:device_classification_performance}(a) Although the method was unable to provide higher rate of classification accuracy on some of the devices but the proposed scheme combined with IP address (as a feature) reached maximum accuracy up to 99.77\%.  
On the other hand, SysID achieved approximately 79.20\% of accuracy using the same dataset. Figure \ref{fig:confusoin_matrix} exhibits the confusion matrix of the proposed algorithm (83.35\% accuracy), while rows define individual device type (or real device) and columns show predicted classification. It would be the optimal case when all the instances from a dataset intersect at the point of both actual and predicted classes. For example, 179 instances from HomeMaticPlug were precisely classified as HomeMaticPlug. However, for the SmarterCoffee, 9 packets were classified correctly as SmarterCoffee and other 7 packets were identified incorrectly as Lightify, which indicated the wrong classification.

\begin{figure*}[ht]
	\centering
	\begin{subfigure}[b]{0.49\textwidth}
		\includegraphics[width=\textwidth, height=9.5cm]{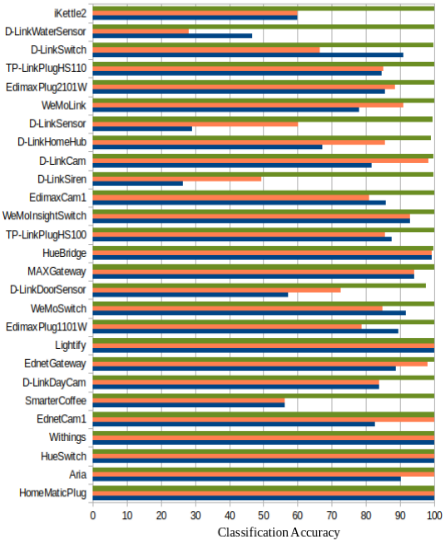}
		\caption{IoT Sentinel dataset}
		\label{fig:devices_27_individual}
	\end{subfigure}
	~ ~ 
	\begin{subfigure}[b]{0.48\textwidth}
		\includegraphics[width=\textwidth, height=9cm]{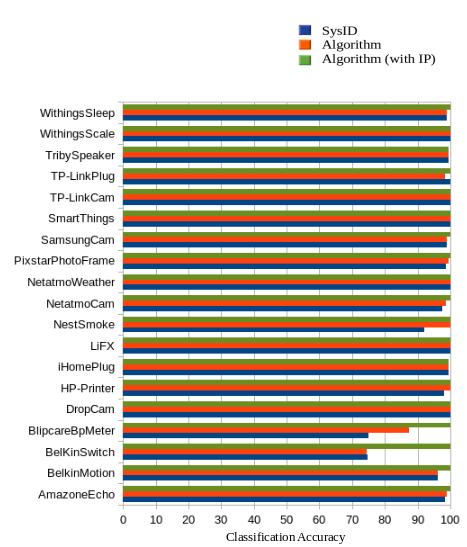}
		\caption{UNSW dataset}
		\label{fig:devices_19_individual}
	\end{subfigure}
	\caption{Individual device type classification performance: (a) IoT Sentinel dataset and (b) UNSW dataset}
	\label{fig:individual_device_classification_performance}
\end{figure*}

Furthermore, Figure \ref{fig:individual_device_classification_performance} (a) exhibits individual device classification accuracy based on the proposed method (with and without IP address) and SysID. Although the proposed scheme (without IP address) classified most of the devices correctly compare with SysID, in some cases like as D-LinkSwitch, TP-LinkPlugHS100, WeMoSwitch, the scheme was not able to identify accurately as SysID. In the experiment on UNSW dataset  \cite{sivanathan2018classifying}, we observed that our proposed method and SysID provided almost similar results, as shown in Figure \ref{fig:device_classification_performance}(b) and Figure \ref{fig:individual_device_classification_performance} (b), where both methods achieved classification accuracy over 97\%. From the both figures, it was noticeable that our proposed method able to achieve 99.93\% of accuracy when we included IP address (such as $ip.src$) as a feature with the selected subset of features. However, Figure \ref{fig:individual_device_classification_performance} (b) presents both methods (SysID and the proposed scheme) performed higher classification precision except for some cases, like as BelkinSwitch and BlipcareBpMeter. From the experimental results on UNSW dataset, we noticed that our proposed DFP method achieved approximately similar accuracy as SysID. Even though both algorithms used a single TCP/IP packet for device identification. SysID performed higher precision when device types varies, but when devices from the same manufacturer performance decline gradually. On the other hand, our proposed DFP scheme exhibits comparatively good result even when a manufacturer with multiple devices.  

Experimental results shown in Figure \ref{fig:individual_device_classification_performance} indicate that the accuracy of the proposed DFP method is improved by using IP address as one of the features to incorporate with the selected feature subset. In a network, every device must have a unique IP address to communicate over the network nevertheless, it is spoofed or altered. Our DFP scheme can utilize this IP address with other features to detect a device type, which helps to mitigate spoofing attack. Hence, our proposed method achieved 99.77\% accuracy from 83.35\% to classify individual device type using IP address as an additional feature on IoT Sentinel dataset. Furthermore tested on UNSW dataset including IP address the proposed DFP approach achieved approximate 99.93\% of accuracy.

\section{Related Work}
The process of gathering device-specific signature from the analysis of network traffic to form a device fingerprint and using them by various classification algorithms to identify individual device type or genre provide significant solutions to distinct research issues. In our study, we mainly focus on network traffic (or packets) analysis based DFP approaches and publicly available commonly used two datasets namely IoT Sentinel \cite{miettinen2017iot} dataset and UNSW \cite{sivanathan2018classifying} dataset, which consists of different types of IoT devices.

A DFP approach proposed based on the analysis of 12 consecutive network packets from a target device by Miettinen et al. \cite{miettinen2017iot}. The authors extracted 23 features from a packet header includes Link layer (ARP, LLC), Network layer (IP, ICMP, ICMPv6, EAPoL), Transport layer (TCP, UDP) and Application layer (HTTP, HTTPS, DHCP, BOOTP, SSDP, DNS, MDNS, NTP) protocols, IP options, Packet content, IP address and Port class to form (12 $\ast$ 23) = 276-dimensional feature vector for a device fingerprint. The set of features represented as binary (0 or 1) and integer values, whereas binary values 1 define some protocols are used during communication. However, from each device, a set of fingerprints are utilized to train a model using Random Forest classification algorithm to perform device identification. The scheme generates one classifier per device type.  While a new device type fingerprints captured by the scheme, it is required to train a new classifier to avoid any modification of the existing classifiers. Overall, the scheme achieved identification accuracy of 81.50\% (global ratio) over the 27 devices from IoT Sentinel dataset \cite{miettinen2017iot}. In comparison, our scheme uses less number of protocol header information from a single packet to construct a device fingerprint and we are able to classify devices (type or genre) by using a single classifier.

Bezawada et al. \cite{bezawada2018iotsense}  presented a device identification mechanism based on the analysis of devices behavioral fingerprint. The researchers extracted 20 features from a packet, which includes distinct protocols from the Link layer, Network layer, Transport layer and Application layer, IP options, entropy of payload, TCP payload length and TCP windows size. While the 17 features are utilized as binary values of 0 or 1 to represent the absence or presence of a particular feature and the 3 other features use as numerical values. The authors grouped device originated 5 packets (as a session) to form a feature vector of (5 $\ast$ 20) = 100 features for a device fingerprinting. Finally, these feature vectors are used to train a machine learning classification algorithm to predict the device type. From the test-bed of 10 unique devices, the scheme achieved a true positive rate of 99.0\% to 100.0\% using all the selected features. Though, the identification accuracy decreased on average 2\% without using payload entropy. In contrast, we use only the packet header information to preserve the privacy of users' data and generate a feature vector from a single TCP/IP packet.

Ammar et al. \cite{ammar2019network} described a DFP approach based on the analysis of service discovery protocols (mDNS and SSDP), DHCP and HTTP protocols on IoT Sentinel \cite{miettinen2017iot} dataset and their own dataset. The authors extracted features as textual information from the first few sets of device originated packets. This textual information from each device modelled as a binary data by using the Bag of Words (BOW) technique for device identification. The scheme first extracts a list of unique words from all devices and then each device information turn into a feature vector, whereas a binary value set to 1 if the word present in the device information and set to 0 if not. Then all the devices information are used to form a matrix while each column defines a unique word and each row represents a device. For classification, devices are evaluated using a similarity verification process. The scheme identified 31 devices among 33 of the unique test devices. However, in some cases, accuracy can be decreased when IoT devices may not use any discovery protocol. Comparing to this scheme, our method uses packet header features of protocols in Table \ref{tab:feature_set} to train a model using ML classification algorithm for device identification.

Guo et al. \cite{guo2018ip} discussed a device identification approach using network traffic analysis and knowledge of devices manufacturer server names. The scheme tracked a list of server names each device communicate regularly and then defined a threshold number of server names require for device identification.  Server names are filtered to exclude third-party and human-facing servers to improve accuracy. The authors examined the presence of server names in network traffic from a given IP address to identify what types of IoT devices present in the network.

From a controlled experiment on 10 IoT devices and 15 non-IoT devices, the reported accuracy of the scheme was 96\%. Our method has some significant differences. We use only packet header features to identify devices without tracking any server names along with IP address. Our scheme can be used to identify IoT devices in a network where manufacturer of the devices are different.  When the manufacturer is same in our scheme e.g. 8 D-Link devices the individual device identification is not sufficiently competent 85.35\% without IP address and  99.77\% with IP address.

Aksoy et al. \cite{aksoy2019automated} proposed a DFP method to identify an IoT device (type or genre) using any single TCP/IP packet content from the device.

The scheme analyzed the Network layer (IP, ICMP), Transport layer (TCP, UDP) and Application layer (DNS, HTTP and SSL) protocols header information to extract 212 features from each packet. Then applied a genetic algorithm (GA) to select feature subsets to create device fingerprint. These selected feature subsets used to train various ML classification algorithms to identify devices. The scheme utilizes two-level of classification to improve accuracy. Whereas first determine the manufacture of devices and then identify individual device type from the same manufacturer using different classifiers. The experiment on IoT Sentinel \cite{miettinen2017iot} dataset over the 23 IoT devices the average classification accuracy was 82\% for the cumulative of all the devices and over 95\% of precision for device genre classification. However, the scheme was not able to achieve a higher rate of accuracy since some devices had similar network behaviors from the same manufacturer. The main difference between our scheme and SysID is feature subset evaluation process. We use character-level analysis of each feature based on the three metrics, while SysID uses a GA to select feature subsets to generate device fingerprinting.

Sivanathan et al. \cite{sivanathan2018classifying} described a framework for IoT devices classification approach by using statistical analysis of network traffic characteristics. The authors extracted network traffic passively to compute traffic characteristics. To characterize network traffic from a device, the authors converted all the raw pcap files into flows using the Joy tool on an hourly basis and then calculated statistical attributes of traffic activity patterns (flow volume, flow duration, average flow rate and device sleep time) and signalling patterns (server port numbers, DNS queries, NTP queries and Cipher suite). This framework consists of two stages, in Stage-0 each multi-valued attributes are fed into corresponding Naive Bayes classifiers in the form of BoW to generate two outputs - a tentative class and a confidence level and outputs are provided into the next stage. In Stage-1, all the quantitative attributes from traffic flow and outputs of Stage-0 use to train a classifier based on the Random Forest algorithm to perform device identification. The accuracy of the experiment was above 99\% for a set of 28 IoT devices. However, the scheme requires a large number of packets from a network flow for statistical analysis. 

In contrast, our scheme uses a single TCP/IP packet header information for classification of devices, while features are evaluated using the three metrics.

Robyns et al. \cite{robyns2017noncooperative} presented mobile devices fingerprinting scheme based on a single Probe-Request frame. The scheme used per bit entropy analysis of the frames from different devices using the three metrics evaluation process to construct device fingerprint (bit patterns), while frames captured by monitoring stations (MSs) without concern of device users. This non-cooperative MAC layer fingerprinting method uses to track mobile devices location. The accuracy of the scheme achieved between 80.0\% to 67.6\% for a small dataset of 50 to 100 devices, while for a large dataset of 1000 to 10000 devices the accuracy varied from 33.0\% to 15.1\%. The scheme provides a mechanism to defeat MAC address randomization problem conjunction with some temporal information. In comparison, our method uses in a home network to capture network packets on an AP and features are extracted from each packet header information to construct suitable feature subset for device fingerprinting. In our DFP scheme, features are evaluated on character-level using the three metrics to identify the most suitable feature subset.

\section{Conclusion}

In this paper, we presented a DFP approach using device originated network traffic (or packets) and machine learning classification algorithms to identify individual device type or genre based on a single packet header features. We learned the most suitable feature subset automatically by utilizing three metrics assessment process of the features. This process assigned a score to every feature according to their significance for device identification, whereas a user-defined threshold value helped to remove noisy features and improve classification accuracy. Overall, the proposed DFP scheme achieved 99.37\% of precision to identify each device genre and 83.35\% of accuracy to classify individual device type from IoT Sentinel dataset. However, for some devices, the proposed method was not able to achieve a higher rate of accuracy since some devices have similar network behaviors from the same manufacturer. On the other hand, the scheme reached a mean accuracy of 97.78\% using UNSW dataset, where most of the devices were from the different manufacturers.

In future, we will investigate other packet header features logged from the network protocols during communication to identify unique feature subset to improve accuracy of the same manufacturer e.g. D-Link 20 devices. Moreover, the problem will be more intense when finding feature for the same device from the same manufacturer e.g. 20 D-Link Cam. 



\bibliographystyle{ACM-Reference-Format}
\bibliography{ref}


\begin{thebibliography}{17}


\ifx \showCODEN    \undefined \def \showCODEN     #1{\unskip}     \fi
\ifx \showDOI      \undefined \def \showDOI       #1{#1}\fi
\ifx \showISBNx    \undefined \def \showISBNx     #1{\unskip}     \fi
\ifx \showISBNxiii \undefined \def \showISBNxiii  #1{\unskip}     \fi
\ifx \showISSN     \undefined \def \showISSN      #1{\unskip}     \fi
\ifx \showLCCN     \undefined \def \showLCCN      #1{\unskip}     \fi
\ifx \shownote     \undefined \def \shownote      #1{#1}          \fi
\ifx \showarticletitle \undefined \def \showarticletitle #1{#1}   \fi
\ifx \showURL      \undefined \def \showURL       {\relax}        \fi
\providecommand\bibfield[2]{#2}
\providecommand\bibinfo[2]{#2}
\providecommand\natexlab[1]{#1}
\providecommand\showeprint[2][]{arXiv:#2}

\bibitem[\protect\citeauthoryear{Aksoy and Gunes}{Aksoy and Gunes}{2019}]%
        {aksoy2019automated}
\bibfield{author}{\bibinfo{person}{Ahmet Aksoy} {and}
  \bibinfo{person}{Mehmet~Hadi Gunes}.} \bibinfo{year}{2019}\natexlab{}.
\newblock \showarticletitle{Automated iot device identification using network
  traffic}. In \bibinfo{booktitle}{\emph{ICC 2019-2019 IEEE International
  Conference on Communications (ICC)}}. IEEE, \bibinfo{pages}{1--7}.
\newblock


\bibitem[\protect\citeauthoryear{Ammar, Noirie, and Tixeuil}{Ammar
  et~al\mbox{.}}{2019}]%
        {ammar2019network}
\bibfield{author}{\bibinfo{person}{Nesrine Ammar}, \bibinfo{person}{Ludovic
  Noirie}, {and} \bibinfo{person}{S{\'e}bastien Tixeuil}.}
  \bibinfo{year}{2019}\natexlab{}.
\newblock \showarticletitle{Network-Protocol-Based IoT Device Identification}.
  In \bibinfo{booktitle}{\emph{2019 Fourth International Conference on Fog and
  Mobile Edge Computing (FMEC)}}. IEEE, \bibinfo{pages}{204--209}.
\newblock


\bibitem[\protect\citeauthoryear{Aneja, Aneja, and Islam}{Aneja
  et~al\mbox{.}}{2018}]%
        {aneja2018iot}
\bibfield{author}{\bibinfo{person}{Sandhya Aneja}, \bibinfo{person}{Nagender
  Aneja}, {and} \bibinfo{person}{Md~Shohidul Islam}.}
  \bibinfo{year}{2018}\natexlab{}.
\newblock \showarticletitle{IoT Device Fingerprint using Deep Learning}. In
  \bibinfo{booktitle}{\emph{2018 IEEE International Conference on Internet of
  Things and Intelligence System (IOTAIS)}}. IEEE, \bibinfo{pages}{174--179}.
\newblock


\bibitem[\protect\citeauthoryear{Bezawada, Bachani, Peterson, Shirazi, Ray, and
  Ray}{Bezawada et~al\mbox{.}}{2018}]%
        {bezawada2018iotsense}
\bibfield{author}{\bibinfo{person}{Bruhadeshwar Bezawada},
  \bibinfo{person}{Maalvika Bachani}, \bibinfo{person}{Jordan Peterson},
  \bibinfo{person}{Hossein Shirazi}, \bibinfo{person}{Indrakshi Ray}, {and}
  \bibinfo{person}{Indrajit Ray}.} \bibinfo{year}{2018}\natexlab{}.
\newblock \showarticletitle{Behavioral fingerprinting of iot devices}. In
  \bibinfo{booktitle}{\emph{Proceedings of the 2018 Workshop on Attacks and
  Solutions in Hardware Security}}. \bibinfo{pages}{41--50}.
\newblock


\bibitem[\protect\citeauthoryear{Douceur}{Douceur}{2002}]%
        {douceur2002sybil}
\bibfield{author}{\bibinfo{person}{John~R Douceur}.}
  \bibinfo{year}{2002}\natexlab{}.
\newblock \showarticletitle{The sybil attack}. In
  \bibinfo{booktitle}{\emph{International workshop on peer-to-peer systems}}.
  Springer, \bibinfo{pages}{251--260}.
\newblock


\bibitem[\protect\citeauthoryear{et~al.}{et~al.}{2018}]%
        {InternetDraft}
\bibfield{author}{\bibinfo{person}{Garcia-Morchon et al.}}
  \bibinfo{year}{2018}\natexlab{}.
\newblock \bibinfo{title}{State-of-the-Art and Challenges for the Internet of
  Things Security draft-irtf-t2trg-iot-seccons-16 (Internet-Draft)}.
\newblock \bibinfo{howpublished}{\url{URL:
  https://tools.ietf.org/pdf/draft-irtf-t2trg-iot-seccons-16.pdf}}.
  (\bibinfo{year}{2018}).
\newblock


\bibitem[\protect\citeauthoryear{Frank, A.~Hall, and H.~Witten}{Frank
  et~al\mbox{.}}{2016}]%
        {Weka}
\bibfield{author}{\bibinfo{person}{Eibe Frank}, \bibinfo{person}{Mark A.~Hall},
  {and} \bibinfo{person}{Ian H.~Witten}.} \bibinfo{year}{2016}\natexlab{}.
\newblock \bibinfo{title}{The WEKA Workbench. Online Appendix for "Data Mining:
  Practical Machine Learning Tools and Techniques", fourth edition ed. Morgan
  Kaufmann, 2016.}
\newblock   (\bibinfo{year}{2016}).
\newblock


\bibitem[\protect\citeauthoryear{Gerald}{Gerald}{1998}]%
        {Gerald1998}
\bibfield{author}{\bibinfo{person}{Combs et~al. Gerald}.}
  \bibinfo{year}{1998}\natexlab{}.
\newblock \bibinfo{title}{D.2. tshark: Terminal-based Wireshark}.
\newblock \bibinfo{howpublished}{\url{URL:
  https://www.wireshark.org/docs/wsug_html_chunked/index.html}}.
  (\bibinfo{year}{1998}).
\newblock
\newblock
\shownote{Accessed on: February 9, 2020.}


\bibitem[\protect\citeauthoryear{Guo and Heidemann}{Guo and Heidemann}{2018}]%
        {guo2018ip}
\bibfield{author}{\bibinfo{person}{Hang Guo} {and} \bibinfo{person}{John
  Heidemann}.} \bibinfo{year}{2018}\natexlab{}.
\newblock \showarticletitle{IP-based IoT device detection}. In
  \bibinfo{booktitle}{\emph{Proceedings of the 2018 Workshop on IoT Security
  and Privacy}}. ACM, \bibinfo{pages}{36--42}.
\newblock


\bibitem[\protect\citeauthoryear{Miettinen, Marchal, Hafeez, Asokan, Sadeghi,
  and Tarkoma}{Miettinen et~al\mbox{.}}{2017}]%
        {miettinen2017iot}
\bibfield{author}{\bibinfo{person}{Markus Miettinen}, \bibinfo{person}{Samuel
  Marchal}, \bibinfo{person}{Ibbad Hafeez}, \bibinfo{person}{N Asokan},
  \bibinfo{person}{Ahmad-Reza Sadeghi}, {and} \bibinfo{person}{Sasu Tarkoma}.}
  \bibinfo{year}{2017}\natexlab{}.
\newblock \showarticletitle{IoT Sentinel: Automated device-type identification
  for security enforcement in IoT}. In \bibinfo{booktitle}{\emph{2017 IEEE 37th
  International Conference on Distributed Computing Systems (ICDCS)}}. IEEE,
  \bibinfo{pages}{2177--2184}.
\newblock


\bibitem[\protect\citeauthoryear{Pinheiro, Bezerra, Burgardt, and
  Campelo}{Pinheiro et~al\mbox{.}}{2019}]%
        {pinheiro2019identifying}
\bibfield{author}{\bibinfo{person}{Ant{\^o}nio~J Pinheiro},
  \bibinfo{person}{Jeandro de~M Bezerra}, \bibinfo{person}{Caio~AP Burgardt},
  {and} \bibinfo{person}{Divanilson~R Campelo}.}
  \bibinfo{year}{2019}\natexlab{}.
\newblock \showarticletitle{Identifying IoT devices and events based on packet
  length from encrypted traffic}.
\newblock \bibinfo{journal}{\emph{Computer Communications}}
  \bibinfo{volume}{144} (\bibinfo{year}{2019}), \bibinfo{pages}{8--17}.
\newblock


\bibitem[\protect\citeauthoryear{Radhakrishnan, Uluagac, and
  Beyah}{Radhakrishnan et~al\mbox{.}}{2014}]%
        {radhakrishnan2014gtid}
\bibfield{author}{\bibinfo{person}{Sakthi~Vignesh Radhakrishnan},
  \bibinfo{person}{A~Selcuk Uluagac}, {and} \bibinfo{person}{Raheem Beyah}.}
  \bibinfo{year}{2014}\natexlab{}.
\newblock \showarticletitle{GTID: A technique for physical deviceanddevice type
  fingerprinting}.
\newblock \bibinfo{journal}{\emph{IEEE Transactions on Dependable and Secure
  Computing}} \bibinfo{volume}{12}, \bibinfo{number}{5} (\bibinfo{year}{2014}),
  \bibinfo{pages}{519--532}.
\newblock


\bibitem[\protect\citeauthoryear{Robyns, Bonn{\'e}, Quax, and Lamotte}{Robyns
  et~al\mbox{.}}{2017}]%
        {robyns2017noncooperative}
\bibfield{author}{\bibinfo{person}{Pieter Robyns}, \bibinfo{person}{Bram
  Bonn{\'e}}, \bibinfo{person}{Peter Quax}, {and} \bibinfo{person}{Wim
  Lamotte}.} \bibinfo{year}{2017}\natexlab{}.
\newblock \showarticletitle{Noncooperative 802.11 mac layer fingerprinting and
  tracking of mobile devices}.
\newblock \bibinfo{journal}{\emph{Security and Communication Networks}}
  \bibinfo{volume}{2017} (\bibinfo{year}{2017}).
\newblock


\bibitem[\protect\citeauthoryear{Shahid and Aneja}{Shahid and Aneja}{2017}]%
        {shahid2017internet}
\bibfield{author}{\bibinfo{person}{Noman Shahid} {and} \bibinfo{person}{Sandhya
  Aneja}.} \bibinfo{year}{2017}\natexlab{}.
\newblock \showarticletitle{Internet of Things: Vision, application areas and
  research challenges}. In \bibinfo{booktitle}{\emph{2017 International
  Conference on I-SMAC (IoT in Social, Mobile, Analytics and Cloud)(I-SMAC)}}.
  IEEE, \bibinfo{pages}{583--587}.
\newblock


\bibitem[\protect\citeauthoryear{Shannon}{Shannon}{1948}]%
        {shannon1948mathematical}
\bibfield{author}{\bibinfo{person}{Claude~E Shannon}.}
  \bibinfo{year}{1948}\natexlab{}.
\newblock \showarticletitle{A mathematical theory of communication}.
\newblock \bibinfo{journal}{\emph{Bell system technical journal}}
  \bibinfo{volume}{27}, \bibinfo{number}{3} (\bibinfo{year}{1948}),
  \bibinfo{pages}{379--423}.
\newblock


\bibitem[\protect\citeauthoryear{Sivanathan, Gharakheili, Loi, Radford,
  Wijenayake, Vishwanath, and Sivaraman}{Sivanathan et~al\mbox{.}}{2018}]%
        {sivanathan2018classifying}
\bibfield{author}{\bibinfo{person}{Arunan Sivanathan},
  \bibinfo{person}{Hassan~Habibi Gharakheili}, \bibinfo{person}{Franco Loi},
  \bibinfo{person}{Adam Radford}, \bibinfo{person}{Chamith Wijenayake},
  \bibinfo{person}{Arun Vishwanath}, {and} \bibinfo{person}{Vijay Sivaraman}.}
  \bibinfo{year}{2018}\natexlab{}.
\newblock \showarticletitle{Classifying IoT devices in smart environments using
  network traffic characteristics}.
\newblock \bibinfo{journal}{\emph{IEEE Transactions on Mobile Computing}}
  \bibinfo{volume}{18}, \bibinfo{number}{8} (\bibinfo{year}{2018}),
  \bibinfo{pages}{1745--1759}.
\newblock


\bibitem[\protect\citeauthoryear{Xu, Zheng, Saad, and Han}{Xu
  et~al\mbox{.}}{2015}]%
        {xu2015device}
\bibfield{author}{\bibinfo{person}{Qiang Xu}, \bibinfo{person}{Rong Zheng},
  \bibinfo{person}{Walid Saad}, {and} \bibinfo{person}{Zhu Han}.}
  \bibinfo{year}{2015}\natexlab{}.
\newblock \showarticletitle{Device fingerprinting in wireless networks:
  Challenges and opportunities}.
\newblock \bibinfo{journal}{\emph{IEEE Communications Surveys \& Tutorials}}
  \bibinfo{volume}{18}, \bibinfo{number}{1} (\bibinfo{year}{2015}),
  \bibinfo{pages}{94--104}.
\newblock


\end{thebibliography}

\appendix








\end{document}